\documentclass[11pt]{article}
\usepackage{cite}  

\begin{document}

\def \d {{\rm d}}
\def \e {e}

\title{Some high-frequency gravitational waves related
to exact radiative spacetimes}

\addvspace{1cm}

\author{
J. Podolsk\'y and O. Sv\' \i tek\footnote{
Institute of Theoretical Physics,
Charles University in Prague,
V~Hole\v{s}ovi\v{c}k\'ach 2, 180~00 Prague 8, Czech Republic}
}
\date{\today}

\maketitle

\begin{abstract}
A formalism  is introduced which may describe both standard linearized
waves and gravitational waves in Isaacson's high-frequency limit.
After emphasizing  main differences between the two approximation techniques
we generalize the Isaacson method  to non-vacuum spacetimes.
Then we present  three large explicit classes of solutions for high-frequency
gravitational waves in particular backgrounds. These involve non-expanding
(plane, spherical or hyperboloidal), cylindrical, and expanding (spherical)
waves propagating in various universes which may contain a cosmological constant
and electromagnetic field. Relations of high-frequency gravitational
perturbations of these types to corresponding exact radiative spacetimes
are described.
\end{abstract}




\section{Introduction}

In classic work \cite{isaac} Isaacson presented a perturbation method which
enables one to study properties of high-frequency gravitational waves, together with
their influence on the cosmological background in which they propagate.
It is this non-linear ``back-reaction'' effect on curvature of the background spacetime
which distinguishes the high-frequency approximation scheme from other perturbation methods
such as the standard Einstein's linearization of gravitational field in flat space
\cite{einstein1,einstein2} or multipole expansions \cite{MTW} that were developed
to describe radiation from realistic astrophysical sources.

The high-frequency perturbations were originally considered by Wheeler \cite{Wheeler} and
then applied to investigation of gravitational geons by Brill and Hartle  \cite{bh}.
Isaacson's systematic study \cite{isaac} stimulated further works in which his treatment
was developed and also re-formulated in various formalisms.
Choquet-Bruhat \cite{choquet1,choquet2}   analyzed
high-frequency gravitational radiation using a generalized WKB ``two-timing'' method.
Averaged Lagrangian technique which leads to Isaacson's results with less calculation
was introduced by MacCallum and Taub \cite{maccallumtaub,taub}. Comparison of
these approaches, and clarification of assumptions that have to be made in order to
provide a consistent high-frequency approximation limit was also given by Araujo
\cite{araujo1,araujo2}. Elster \cite{elster} proposed an alternative method
that is based on expanding null-tetrad components of the Weyl tensor.
Recently,  Burnett developed a  weak limit approach \cite{burnett}
in which the high-frequency limit can be introduced and studied in a mathematically rigorous way.
These general methods have been, of course,  applied to study explicit particular
examples of high-frequency gravitational waves, see e.g.
\cite{isaac,choquet2,maccallumtaub,taub2,hogan}.

On the other hand, many  \emph{exact} solutions of Einstein's equations
are known which represent gravitational radiation. Among  the most important
classes are planar {\it pp\,}-waves \cite{brinkmann,BPR} which belong to a large family
of non-expanding radiative spacetimes \cite{kundt,ehlers}, cylindrical
Einstein-Rosen waves \cite{einst-ros}, expanding ``spherical'' waves of the Robinson-Trautman
type \cite{rob-traut1,rob-traut2}, spacetimes with boost-rotation symmetry representing
radiation generated by uniformly accelerated sources \cite{BonSwa64,Bic68,KinWal70},
cosmological models of the Gowdy type \cite{gowdy}, and  others --- for
comprehensive reviews containing also a number of references see, e.g.,
\cite{kramerbook,BicSch89b,BoGrMa94,carmeli-charach,Bic}.

However, there are only several works in which \emph{relation} between
exact gravitational waves and those obtained by perturbations of non-flat backgrounds
has been explicitly investigated and clarified, see e.g. \cite{taub,feinstein88,hogan}.
The purpose of our contribution is to help to fill this ``gap''.

We first briefly summarize and generalize the Isaacson approach \cite{isaac} to admit
non-vacuum backgrounds, the cosmological constant $\Lambda$ in particular. Modification
of Isaacson's formalism allows us to incorporate also standard linearized gravitational waves
into the common formalism. Then, in section~3 we study properties
of high-frequency gravitational waves in specific classes of spacetimes with special
algebraic or geometric structure. In particular, we focus on waves which
propagate in backgrounds with ${\Lambda\not=0}$. This is motivated
not only theoretically but also by recent observations \cite{WMAP} which
seem to indicate that (effective) positive cosmological constant played a fundamental role
in the early universe, but it is also important for its
present and future dynamics.


\section{High-frequency approximation versus standard linearization}

Let us assume a formal decomposition of the spacetime metric $g_{\mu\nu}$ into the background metric
$\gamma_{\mu\nu}$  and its perturbation $h_{\mu\nu},$
\begin{equation}\label{x1.1}
g_{\mu\nu}=\gamma_{\mu\nu}+\varepsilon h_{\mu\nu}\ ,
\end{equation}
where, in a suitable coordinate system, ${\gamma_{\mu\nu}=O(1)}$ and
${h_{\mu\nu}=O(\epsilon)}$
[by definition, $f=O(\epsilon^n)$ if there exists a constant $C>0$ such that
$|f|<C\epsilon^n$ as $\epsilon\to 0$].
The two distinct non-negative dimensionless parameters $\varepsilon$ and $\epsilon$ have the following
meaning: $\varepsilon$ is the usual amplitude parameter of weak gravitational perturbations
whereas the frequency parameter $\epsilon$ denotes the possible high-frequency character of
radiation described by $h_{\mu\nu}$.
To be more specific, the  parameter ${\varepsilon\ll 1}$ characterizes  (for ${\epsilon=1}$)
the amplitude of linearized gravitational waves in the ordinary weak field limit of Einstein's
equations. The second independent parameter ${\epsilon=\lambda /L}$
represents, on the other hand, the ratio of a typical wavelength $\lambda$ of gravitational waves
and the scale $L$ on which the background curvature changes significantly. Isaacson's high-frequency
approximation  \cite{isaac} arises when ${\lambda\ll L}$, i.e. ${\epsilon\ll 1}$ (and ${\varepsilon=1}$).
Since  $L$ can be considered to have a finite value of order unity, we may write
 ${O(\epsilon)=O(\lambda)}$.

To derive the dynamical field equations we start with the order-of-magnitude estimates which indicate
how fast the metric components vary. Symbolically, the derivatives
are of the order ${\partial\gamma\sim\gamma/L}$, ${{\partial}h\sim h/\lambda}$, so that  the following formulas
\begin{equation}\label{x1.4}
\begin{array}{rclcrcl}
\gamma_{\mu\nu}&=&O(1)\ ,&&h_{\mu\nu}&=&O(\epsilon)\ ,\\
\gamma_{\mu\nu ,\alpha}&=&O(1)\ ,&&h_{\mu\nu ,\alpha}&=&O(1)\ ,\\
\gamma_{\mu\nu ,\alpha\beta}&=&O(1)\ ,&&h_{\mu\nu ,\alpha\beta}&=&O({\epsilon}^{-1})\ ,
\end{array}
\end{equation}
are valid. Next, we expand the Ricci tensor in powers of $h$,
\begin{equation}\label{2.5}
R_{\mu\nu}(g)=R^{(0)}_{\mu\nu}+{\varepsilon}R^{(1)}_{\mu\nu}+{\varepsilon}^2 R^{(2)}_{\mu\nu}
+{\varepsilon}^3 R^{(3)}_{\mu\nu}+\ldots\ ,
\end{equation}
where
\begin{eqnarray}
R^{(0)}_{\mu\nu}(\gamma)&\equiv &R_{\mu\nu}(\gamma)\ ,\nonumber \\ 
R^{(1)}_{\mu\nu}(\gamma ,h)&\equiv &{\textstyle \frac{1}{2}}\gamma^{\rho\tau}
   \left(h_{\tau\mu ;\nu\rho}+h_{\tau\nu ;\mu\rho}
   -h_{\rho\tau ;\mu\nu}-h_{\mu\nu ;\rho\tau}\right)\ ,\label{2.7}\\
R^{(2)}_{\mu\nu}(\gamma ,h)&\equiv &{\textstyle  \frac{1}{2}}
   \Big[{\textstyle \frac{1}{2}}h^{\rho\tau}{}_{;\nu}h_{\rho\tau ;\mu}+
   h^{\rho\tau} (h_{\tau\rho ;\mu\nu}+h_{\mu\nu
   ;\tau\rho}-h_{\tau\mu;\nu\rho}   \nonumber\\
&&  \quad - h_{\tau\nu ;\mu\rho}) +  h^{\tau}{}_{\nu}{}^{;\rho}\left( h_{\tau\mu ;\rho}-h_{\rho\mu;\tau}\right)\nonumber\\
&&  \quad -\left(h^{\rho\tau}{}_{;\rho}-{\textstyle \frac{1}{2}}h^{;\tau}\right)
   \left(h_{\tau\mu ;\nu}+h_{\tau\nu ;\mu}-h_{\mu\nu ;\tau}
   \right)\Big]\; .\nonumber\\
R^{(3)}_{\mu\nu}(\gamma ,h)&\equiv &{\textstyle \frac{1}{4}}h^{\sigma\tau}h_{\sigma\rho ;\mu}h^{\rho}{}_{\tau ;\nu}+\ldots\ . \nonumber
\end{eqnarray}
The semicolons denote covariant differentiation with respect to the \emph{background} metric $\gamma_{\mu\nu}$,
which is also used to raise or lower all indices. Considering  relations (\ref{x1.4}),
the orders of the terms (\ref{2.7}) are
\begin{equation}\label{rexp}
R^{(0)}_{\mu\nu}=O(1),\
{\varepsilon}R^{(1)}_{\mu\nu}=O({\epsilon}^{-1}{\varepsilon}),\
{\varepsilon}^2 R^{(2)}_{\mu\nu}=O({\varepsilon}^2),\ {\varepsilon}^3 R^{(3)}_{\mu\nu}=O(\epsilon {\varepsilon}^3).
\end{equation}

Two limiting cases thus arise naturally.
For the \emph{standard linearization} ({$\varepsilon \ll 1$}, {$\epsilon=1$}) the dominant term
of $R_{\mu\nu}(g)$ is ${R^{(0)}_{\mu\nu}=O(1)}$
which corresponds to the background $\gamma_{\mu\nu}$ [to find, e.g., a vacuum spacetime metric $g_{\mu\nu}$
we solve  ${R^{(0)}_{\mu\nu}(\gamma)=0}$]. Its first correction representing  linearized
(purely) gravitational waves is governed by
\begin{equation}\label{R1}
R^{(1)}_{\mu\nu}(\gamma ,h)=0\ ,
\end{equation}
which is a dynamical equation for perturbations $h_{\mu\nu}$ on the fixed background $\gamma_{\mu\nu}$.
The next term ${R^{(2)}_{\mu\nu}(\gamma ,h)}$ can then be used to define  energy-momentum tensor of these
gravitational waves, but the background metric is \emph{not} assumed to be influenced by it. Improvements
to this inconsistency can be obtained by iteration procedure. More rigorous but somewhat complicated
solution to this problem was recently proposed by Efroimsky {\cite{efroimsky}}.

In the \emph{high-frequency approximation} (${\epsilon \ll 1}$, ${\varepsilon=1}$) the dominant
term is $R^{(1)}_{\mu\nu}=O({\epsilon}^{-1})$ which gives the wave equation (\ref{R1})
for the perturbations $h_{\mu\nu}$ on the curved background $\gamma_{\mu\nu}$
(considering a vacuum full metric $g_{\mu\nu}$).
The two terms of the order $O(1)$, namely  $R^{(0)}_{\mu\nu}$ and $R^{(2)}_{\mu\nu}$, are \emph{both} used
to give the Einstein equation for the background \emph{non-vacuum} metric, which represents the essential
influence of the high-frequency gravitational waves on the background. Of course, to obtain a consistent
solution, one has to use both the wave equation \emph{and} the Einstein equation for the background simultaneously.

\subsection{Linear approximation}

Interestingly, it follows that the wave equation for $h_{\mu\nu}$, which arises from the linear
perturbation of the Ricci tensor in vacuum for \emph{both} the above  limiting cases
${\varepsilon \ll 1}$, {$\epsilon=1$}, and {$\epsilon \ll 1$},
${\varepsilon=1}$, is the \emph{same} equation (\ref{R1}). In analogy with the well-known theory of massless
spin-2 fields in flat space \cite{MTW} we wish to impose two TT gauge
conditions,
\begin{eqnarray}
{h_{\mu\nu}}^{;\nu}&=&0\ ,\label{x1.21}\\
h^\mu_{\ \mu} &=&0\ . \label{x1.22}
\end{eqnarray}
In this gauge we arrive at the following wave equation
\begin{equation}\label{x1.26}
\diamondsuit h_{\mu\nu}\equiv{{h_{\mu\nu}}^{;\beta}}_{;\beta}-
2R^{(0)}_{\sigma\nu\mu\beta}\,{h}^{\beta\sigma}-R^{(0)}_{\mu\sigma}\,
{h^{\sigma}}_{\nu}-R^{(0)}_{\nu\sigma}\,{h^{\sigma}}_{\mu}=0\ ,
\end{equation}
where the operator $\diamondsuit$ is the generalization of flat-space d'Alembertian.
Contracting (\ref{x1.26}) we obtain ${{(h^\mu_{\ \mu})}^{;\beta}}_{;\beta}=0$, so that
the condition  (\ref{x1.22}) is always consistent with (\ref{x1.26}).
However, if we differentiate $\diamondsuit h_{\mu\nu}$ and use equations (\ref{x1.21}),
(\ref{x1.4}), we find that
\begin{eqnarray}\label{e6}
&&(\diamondsuit h_{\mu\nu})^{;\nu}=(R^{(0)}_{\nu\beta ;\mu}-2R^{(0)}_{\mu\nu ;\beta})h^{\nu\beta}\ ,\quad
\hbox{where}\\
&&(\diamondsuit h_{\mu\nu})^{;\nu}=O({\epsilon}^{-2})\ ,\quad (R^{(0)}_{\nu\beta ;\mu}-2R^{(0)}_{\mu\nu ;\beta})h^{\nu\beta}=O(\epsilon)\  .
\end{eqnarray}
Thus, in case of standard linearized waves (${\epsilon=1}$) there is an obvious inconsistency,
except for backgrounds with a covariantly constant Ricci tensor (e.g., for the Einstein spaces).
On the other hand, in the high-frequency limit (${\varepsilon=1}$), the inconsistency between (\ref{x1.26})
and (\ref{x1.21}) is extremely small (the left and the right sides of (\ref{e6}) differ by
${\epsilon^3}$ where ${\epsilon\ll1}$). Moreover, for all background metrics of \emph{constant curvature}
the equations are \emph{fully consistent}. This is an important advantage of the
equation (\ref{x1.26}) containing also terms of non-dominant order (namely those
proportional to the Riemann or Ricci tensors),
if compared to other ``simpler" wave equations (e.g., ${h_{\mu\nu}{}^{;\beta}{}_{;\beta}=0}$)
for which the left and right sides of (\ref{e6}) generally differ by only two orders of magnitude.

\subsection{Nonlinear  terms and the effective energy-momentum tensor}

Before considering the second-order terms  we now extend the formalism to be applicable to a larger
class of spacetimes with (possibly) non-vanishing energy-momentum tensor $T_{\mu\nu}$. Namely,
$g_{\mu\nu}$ need not be a vacuum metric (as only considered in \cite{isaac}) but it satisfies
Einstein's equations
\begin{equation}\label{my.1}
R_{\mu\nu}(g)=8\pi\, \tilde{T}_{\mu\nu}(g,\varphi)\ .
\end{equation}
Here ${\tilde{T}_{\mu\nu}\equiv T_{\mu\nu}-{\textstyle \frac{1}{2}}g_{\mu\nu}{T^\beta}_\beta}$,
such that $T_{\mu\nu}(g,\varphi)$ depends on non-gravitational fields $\varphi$ and on the full metric
$g_{\mu\nu }$ but it \emph{does not} contain the \emph{derivatives} of $g_{\mu\nu}$. Note
that this admits as particular cases a presence of electromagnetic field, and also Einstein
spaces when ${\tilde{T}_{\mu\nu}=\frac{1}{8\pi}\,\Lambda g_{\mu\nu}}$. Under the assumptions
(\ref{x1.4}) valid for the decomposition (\ref{x1.1}) we expand the equation (\ref{my.1}) as
\begin{eqnarray}
&&R^{(0)}_{\mu\nu}(\gamma)+\varepsilon R^{(1)}_{\mu\nu}(\gamma ,h)
   +\varepsilon^2R^{(2)}_{\mu\nu}(\gamma ,h)+\ldots=\label{my.2}\\
&&\qquad\qquad 8\pi\, [\,\tilde{T}^{(0)}_{\mu\nu}(\gamma,\varphi)+\varepsilon\, \tilde{T}^{(1)}_{\mu\nu}(\gamma,h,\varphi)
   +\varepsilon^2\,\tilde{T}^{(2)}_{\mu\nu}(\gamma,h,\varphi)+\ldots\,]\ ,\nonumber
\end{eqnarray}
where
$\tilde{T}^{(0)}_{\mu\nu}(\gamma,\varphi)\equiv \tilde{T}_{\mu\nu}(\gamma,\varphi),$ and
 the remaining terms on the right-hand side are linear and quadratic in $h$, respectively.
The orders of magnitude of the terms in the expansion of the Ricci tensor have been  described above,
cf. (\ref{rexp}).
For the energy-momentum tensor one obtains
\begin{equation}
\tilde{T}^{(0)}_{\mu\nu}=O(1),\quad \tilde{T}^{(1)}_{\mu\nu}=O(\epsilon)
,\quad \tilde{T}^{(2)}_{\mu\nu}=O(\epsilon^{2}) .
\end{equation}
For ordinary linearization we thus get the equations
${R^{(n)}_{\mu\nu}= 8\pi\, \tilde{T}^{(n)}_{\mu\nu}}$ in each order ${n=0,1,2,\ldots}$.
For the high-frequency approximation we obtain from (\ref{my.2}) in the leading order
$O(\epsilon^{-1})$ the equation (\ref{R1})
which is identical with the wave equation in the vacuum case.
The second-order contributions, that are $O(1)$, represent an influence of the
high-frequency gravitational waves and matter fields on the background,
\begin{equation}\label{my.4}
R_{\mu\nu}^{(0)}(\gamma)-8\pi \tilde{T}^{(0)}_{\mu\nu}(\gamma,\varphi)=
-R_{\mu\nu}^{(2)}(\gamma ,h)\ .
\end{equation}
This equation (which in case of a vacuum spacetime reduces to the Isaacson result) can be rewritten in the form
of Einstein's equation for the background as
\begin{equation}\label{my.5}
G^{(0)}_{\mu\nu}(\gamma)-8\pi\, T^{(0)}_{\mu\nu}(\gamma ,\varphi)=-
[R_{\mu\nu}^{(2)}(\gamma ,h)-{\textstyle \frac{1}{2}}\gamma_{\mu\nu}
R^{(2)}(\gamma ,h)]\equiv 8\pi\, T^{GW}_{\mu\nu}\ .
\end{equation}
This defines the effective energy-momentum tensor $T^{GW}_{\mu\nu}$ of
high-frequency gravitational waves.

\subsection{Gravitational waves in the WKB approximation}

In the following we shall restrict ourselves to the Isaacson approximation (${\varepsilon=1}$, ${\epsilon\ll 1}$),
i.e. on study of high-frequency gravitational waves on curved backgrounds. Inspired by the
plane-wave solution in flat space, the  form ${h_{\mu\nu}=\mathcal{A}\,e_{\mu\nu}\exp({i\phi})}$ of the solution
is assumed. The amplitude ${\mathcal{A}=O(\epsilon)}$ is a slowly changing real
function of position, the phase $\phi$ is a real function with a large first
derivative but no larger derivatives beyond, and $e_{\mu\nu}$ is a normalized
polarisation tensor field. The above assumption, introduced in \cite{isaac},
is called the WKB approximation, or the geometric optics limit \cite{MTW}.
The wave vector normal to surfaces of constant phase is
$k_{\mu}\equiv \phi_{,\mu}$
and the orders of various relevant quantities are
${R^{(0)}_{\mu\nu\gamma\delta}= O(1)}$,
${\mathcal{A}_{,\mu}=O(\epsilon)}$,
${k_{\mu}=O({\epsilon}^{-1})}$,
and ${k_{\mu ;\nu}=O({\epsilon}^{-1})}$.
Substituting this into the conditions (\ref{x1.21}), (\ref{x1.22}),
and the wave equation (\ref{x1.26}) we obtain, in the two highest orders which are gauge invariant,
\begin{eqnarray}\label{2.4.4b}
&&k^{\mu}k_{\mu}=0\ ,\quad
  k^{\mu}e_{\mu\nu}=0\ ,\quad
  k^{\alpha}e_{\mu\nu ;\alpha}=0\ ,\nonumber\\
&&e^{\mu\nu}e_{\mu\nu}=1\ ,\quad
  \gamma^{\mu\nu}e_{\mu\nu}=0\ ,\quad
  \left(\mathcal{A}^{2}k^{\beta}\right)_{;\beta}=0\ .
\end{eqnarray}
These express that a beam of high-frequency gravitational waves propagate along
rays which are null geodesics with tangent $k^\mu$, with parallelly transported
polarization orthogonal to the rays. Moreover, using the  WKB approximation of
$T^{GW}_{\mu\nu}$ and the Brill-Hartle averaging procedure \cite{bh}
(which guarantees the gauge invariance) Isaacson  obtained for gravitational waves in the
geometric optics limit the energy-momentum tensor \cite{isaac}
\begin{equation}\label{2.4.6}
T^{HF}_{\mu\nu}={\textstyle \frac{1}{64\pi}}\mathcal{A}^{2}k_{\mu}k_{\nu}\ .
\end{equation}
The  energy-momentum tensor of high-frequency  waves thus has the form of pure radiation.
This fully agrees with  results obtained by alternative
techniques \cite{choquet2,maccallumtaub,burnett}.

\section{Examples of high-frequency gravitational waves}

Now we present some explicit classes of high-frequency gravitational waves. These are obtained
by the above described WKB approximation method considering specific families of
background spacetimes with a privileged geometry.

\subsection{Non-expanding waves}
As the background we first consider the Kundt class \cite{kundt,kramerbook} of non-expanding,
twist-free spacetimes in the form \cite{PodOrt03}
\begin{equation}
 \d s^2=F\,\d  u^2-2\,{Q^2\over P^2}\,\d u\,\d v+{1\over P^2}\,(\d x^2+\d y^2)\ ,
 \label{kundt}
\end{equation}
with
\begin{eqnarray}
 &&P = 1+\frac{\alpha}{2}\,(x^2+y^2)\ , \nonumber\\
 &&Q = \Big[1+\frac{\beta}{2}\,(x^2+y^2)\Big]\,\e+C_1\,x+C_2\,y\ ,  \label{coeff} \\
 &&F = D\,{Q^2\over P^2}\,v^2-\frac{(Q^2)_{,u}}{P^2}\,v-\frac{Q}{P}\,H\ , \nonumber
\end{eqnarray}
where $\alpha$, $\beta$, and $\e$ are constants (without loss of generality
$\e=0$ or $\e=1$), $C_1$, $C_2$ and $D$ are arbitrary functions of the retarded time $u$, and
$H(x,y,u)$ is an arbitrary function of the spatial
coordinates $x$, $y$, and of $\,u$.

In particular, these are Petrov type~$N$ (or conformally flat) solutions of
Einstein's  equations with cosmological constant $\Lambda$
 when ${\alpha=-\beta=\frac{1}{6}}\Lambda\,$ and
${D=-2\beta\e+C_1^2+C_2^2}$, see e.g. \cite{OzsRobRoz85,Siklos85,BicPod99I,PodOrt03}.
Such metrics represent exact pure gravitational waves
propagating  along  principal null direction $\partial_v$ if
$H$ satisfies the equation $P^2(H_{,xx}+H_{,yy})+\frac{2}{3}\Lambda\,H=0$.
However, in our treatment here the function $H$ does \emph{no}t describe
exact gravitational waves but rather it characterizes the \emph{influence}
of high-frequency perturbations on the background metric, which is assumed
to be initially given by (\ref{kundt}), (\ref{coeff}) with ${H=0}$.

We consider the phase of high-frequency gravitational waves given by
$\phi=\phi(u)$, and we seek solution in the WKB form, namely
\begin{equation}\label{x1.33b}
h_{\mu\nu}=\mathcal{A}\, e_{\mu\nu}\exp\Big({i\phi(u)}\Big)\ ,
\end{equation}
where the amplitude $\mathcal{A}$ and  polarization tensor $e_{\mu\nu}$ are functions
of the coordinates ${\{u,v,x,y\}}$. The corresponding wave vector is
${k_{\mu}=(\dot{\phi},0,0,0)}$, where the dot denotes differentiation with respect to $u$.
Applying now all the equations (\ref{2.4.4b}) we obtain
\begin{eqnarray}
\mathcal{A}&=&\mathcal{A}(u,x,y)\ ,  \nonumber\\
e^{+}_{\mu\nu}&=&\frac{1}{\sqrt{2}\,P^{2}}
\left(
\begin{array}{cccc}
0&0&0&0\\
0&0&0&0\\
0&0&1&0\\
0&0&0&-1
\end{array}
\right)\ ,\label{tenz}\\
e^{\times}_{\mu\nu}&=&\frac{1}{\sqrt{2}\,P^{2}}
\left(
\begin{array}{cccc}
0&0&0&0\\
0&0&0&0\\
0&0&0&1\\
0&0&1&0
\end{array}
\right)\ .\nonumber
\end{eqnarray}
The fact that the amplitude ${\cal A}$ is independent of the
coordinate $v$ expresses  non-expanding character of the waves.
The special polarisation tensors, denoted as $+$ and $\times$, are analogous to those
used in the standard theory of linearized waves in flat space. A general polarisation
 is easily obtained by considering ${e_{\mu\nu}=a\,e^{+}_{\mu\nu}+b\,e^{\times}_{\mu\nu}}$, where
${a^{2}(u,x,y)+b^{2}(u,x,y)=1}$.

Using the Einstein tensor for the metric (\ref{kundt}) with the cosmological term in equations (\ref{my.5})
and (\ref{2.4.6}), we determine the reaction of the background on the presence of the above high-frequency
gravitational perturbations, namely
\begin{equation}\label{rovback}
\frac{Q}{P}\left[P^2\left(\frac{\partial^{2}}{\partial x^2}+\frac{\partial^{2}}{\partial y^2}\right)+\frac{2}{3}\,\Lambda\right]H(u,x,y)=
{\textstyle \frac{1}{4}}\mathcal{A}^{2}(u,x,y)\dot{\phi}^{2}\ .
\end{equation}
Notice that ${\mathcal{A}=O(\epsilon)}$ and ${\dot{\phi}=O({\epsilon}^{-1})}$. Therefore, the influence
of high-frequency gravitational waves on the background,  represented by the function $H$, is of
the order  $O(1)$. These \emph{approximate} solutions can obviously be compared to specific \emph{exact}
radiative vacuum solutions which are given by $H$  solving the field equation (\ref{rovback}) with
a vanishing right-hand side (when ${\mathcal{A}=0}$, i.e. high-frequency perturbation waves are absent).

The above waves are non-expanding with the wave-fronts ${u=const.}$ being two-dimensional spaces of constant curvature given by ${\alpha=\frac{1}{6}}\Lambda\,$, cf. (\ref{kundt}). For ${\Lambda=0}$ these are
plane-fronted waves, for ${\Lambda>0}$ they are spheres, and for ${\Lambda<0}$ hyperboloidal surfaces.

Another interesting subclass of the Kundt spacetimes of the form (\ref{kundt}),
(\ref{coeff}) are explicit Petrov type~$II$ (or more special) metrics
given by $\beta=\alpha$, $\e=1$, $C=0$ and
$D=2(\Lambda-\alpha)$, namely
\begin{equation}\label{kundtii}
ds^{2}=\Big[\,2(\Lambda-\alpha)\,v^{2}-H\,\Big]\,\d u^{2}-2\,\d u\,\d v+{1\over P^2}\,(\d x^2+\d y^2)\ .
\end{equation}
For $H=0$ these are electrovacuum solutions with the geometry of a
direct product of two 2-spaces of constant curvature, in particular
the Bertotti-Robinson, (anti-)Nariai or Pleba\'nski-Hacyan spaces
\cite{Bertotti59,Robinson59,Nariai51,PlebHac79}, see e.g. \cite{OrtPod02,PodOrt03}.
Considering again (\ref{x1.33b}) we obtain the results (\ref{tenz}) as in the previous case.
However, the reaction of high-frequency waves on the background is now different. It is determined
by the equations (\ref{my.5}) and (\ref{2.4.6}) with the energy-momentum tensor consisting of a
cosmological term plus that of a uniform non-null electromagnetic field described by
the complex self-dual Maxwell tensor $F^{\mu\nu}=4\Phi_1(m^{[\mu}\bar{m}^{\nu]}-k^{[\mu}l^{\nu]})$,
where $\Phi_{1}=\sqrt{\alpha-\frac{\Lambda}{2}}\,e^{i\,c}$, $c=const.$, and
${{\bf m}= P\,\partial_{\bar{\zeta}}}$,
${{\bf k}=\partial_v}$, ${{\bf l}={1\over 2}F\,\partial_v+\partial_u}$ form the null tetrad.
Straightforward calculation gives
\begin{equation}\label{reac}
P^{2}\left(\frac{\partial^{2}}{\partial x^2}+\frac{\partial^{2}}{\partial y^2}\right)\,H
={\textstyle \frac{1}{4}}\mathcal{A}^{2}(u,x,y)\dot{\phi}^{2}\ .
\end{equation}

This result is  analogous to the equation (\ref{rovback}), but the present situation is now more complicated since
the background spacetime is \emph{not vacuum} but it contains electromagnetic field.
(In fact, the term with the cosmological constant $\Lambda$ in (\ref{rovback}) has been entirely compensated by this.)
Therefore, we have to analyze the perturbation of the \emph{complete} Einstein-Maxwell system, and its consistency.

The Einstein  equations in the two highest orders (\ref{R1}) and (\ref{my.5}) have already been solved.
We will now demonstrate that the Maxwell equations are also satisfied
in the high-frequency limit, namely $F^{\mu\nu}{}_{|\nu}=O(\epsilon)$, where $|$ denotes the covariant derivative
with respect to the full metric $g_{\mu\nu}$.
Indeed, using antisymmetry of $F^{\mu\nu}$ we can write
$F^{\mu\nu}{}_{|\nu}=F^{\mu\nu}{}_{,\,\nu}+{\textstyle \frac{1}{2}} g^{\alpha\beta} g_{\alpha\beta ,\nu}F^{\mu\nu}$.
Considering (\ref{x1.4}) and the gauge condition (\ref{x1.22}) we obtain
${g^{\alpha\beta} g_{\alpha\beta ,\nu}=\gamma^{\alpha\beta} \gamma_{\alpha\beta ,\nu}
-h^{\alpha\beta} h_{\alpha\beta ,\nu}+O(\epsilon^2)}$ because
${\gamma^{\alpha\beta} h_{\alpha\beta ,\nu}-h^{\alpha\beta} \gamma_{\alpha\beta ,\nu}
={{(h^\beta_{\ \beta})}}_{;\nu}-2\,h^{\alpha\beta} \gamma_{\alpha\beta ;\nu}=0\,}$, so that
\begin{equation}\label{Max}
F^{\mu\nu}{}_{|\nu}=F^{\mu\nu}{}_{;\nu}
  -{\textstyle \frac{1}{2}} h^{\alpha\beta} h_{\alpha\beta ,\nu}F^{\mu\nu}+O(\epsilon^2)\ .
\end{equation}
Consequently, if the original background represents an electrovacuum spacetime, ${F^{\mu\nu}{}_{;\nu}=0}$,
the Maxwell equations ${F^{\mu\nu}{}_{|\nu}=O(\epsilon)}$ for the
full metric are satisfied in the dominant order $O(1)$ in the high-frequency limit ${\epsilon\ll1}$.
In addition, the field equations are valid also in the next order
$O(\epsilon)$ for the new electromagnetic  field
\begin{equation}\label{newfield}
{\cal F}^{\mu\nu}=(1+{\textstyle \frac{1}{4}} h^{\alpha\beta} h_{\alpha\beta})\,F^{\mu\nu}\ ,
\end{equation}
since using (\ref{Max}) we obtain ${{\cal
F}^{\mu\nu}{}_{|\nu}=O(\epsilon^2)}$. Starting from an electromagnetic field $F^{\mu\nu}$
satisfying ${F^{\mu\nu}{}_{;\nu}=0}$ with respect to the background metric ${\gamma_{\mu\nu}}$,
we can thus construct the electromagnetic field ${\cal F}^{\mu\nu}$ which satisfies the Maxwell equations
${\cal F}^{\mu\nu}{}_{|\nu}=O(\epsilon^2)$ with respect to the full metric $g_{\mu\nu}$ in the
presence of high-frequency gravitational waves. Both the Einstein and Maxwell equations
are then satisfied in the two highest perturbative orders. Interestingly, these results hold for
high-frequency perturbations of \emph{any} ``seed'' electrovacuum background spacetimes.

In particular, if the backgrounds are direct product spacetimes
(\ref{kundtii}) for ${H=0}$ with uniform non-null electromagnetic
field $\Phi_1=const.$ then high-frequency gravitational waves
(\ref{x1.33b}), (\ref{tenz}) introduce $H$ which is given by equation (\ref{reac}).
According to (\ref{newfield}), the electromagnetic field  is perturbed
by the term proportional to
$h^{\alpha\beta}h_{\alpha\beta}=\mathcal{A}^2e^{2i\phi}=O(\epsilon^2)$, see (\ref{2.4.4b}),
namely
\begin{equation}
\Phi^g_{1}=\Phi_{1}\left[1+{\textstyle
\frac{1}{2}}\mathcal{A}^{2}(u,x,y)\,e^{2i\phi(u)}\right]\ .
\end{equation}
This remains non-null but it is no longer uniform. The full spacetime thus
describes non-uniform, non-null electromagnetic field plus the
null  field of high-frequency gravitational waves.

\subsection{Cylindrical waves}
Next we consider the class of cylindrical Einstein-Rosen waves,
\begin{equation}
ds^{2}=e^{2\gamma-2\psi}(-\d t^{2}+\d\rho^{2})+
e^{2\psi}\d z^{2}+\rho^{2}e^{-2\psi}\d\varphi^{2}\ .\label{cylmetrp}
\end{equation}
If the functions $\psi(t,\rho)$ and $\gamma(t,\rho)$ satisfy the corresponding field equations
(see, e.g. \cite{einst-ros},\cite{kramerbook}, or
equations (\ref{r1})-(\ref{r3}) below) these are exact radiative spacetimes of the Petrov type~I.
We conveniently define double null coordinates ${u=\frac{1}{\sqrt{2}}(t-\rho)}$ and
${v=\frac{1}{\sqrt{2}}(t+\rho)}$; in these coordinates ${\{u,v,\varphi ,z\}}$ the  metric takes the  form
\begin{equation}
ds^{2}=-2\,e^{2\gamma-2\psi}\d u\,\d v+e^{2\psi}\d z^{2}+{\textstyle\frac{1}{2}}(v-u)^{2}e^{-2\psi}\d\varphi^{2}\ .
\label{cylmetr}
\end{equation}
We assume this to be the class of background universes into which we wish to introduce high-frequency
gravitational waves. We assume again $\phi=\phi(u)$ implying the wave vector
${k_{\mu}=(\dot{\phi},0,0,0)}$, i.e. the WKB perturbation of the form (\ref{x1.33b}).
By applying all the conditions (\ref{2.4.4b}) we obtain
\begin{eqnarray}
\mathcal{A}&=&\frac{\mathcal{U}(u)}{\sqrt{v-u}}\ ,\nonumber\\
e^{+}_{\mu\nu}&=&\frac{1}{\sqrt{2}}\,e^{-2\psi}
\left(
\begin{array}{cccc}
0&0&0&0\\
0&0&0&0\\
0&0&\frac{1}{2}(v-u)^{2}&0\\
0&0&0&-e^{4\psi}
\end{array}
\right)\ ,\label{amplit}\\
e^{\times}_{\mu\nu}&=&\frac{1}{2}\,(v-u)
\left(
\begin{array}{cccc}
0&0&0&0\\
0&0&0&0\\
0&0&0&1\\
0&0&1&0
\end{array}
\right)\ ;\nonumber
\end{eqnarray}
notice that ${v-u=\sqrt{2}\,\rho>0}$. Thus the perturbative solution
is given by
\begin{equation}
h_{\mu\nu}=\frac{\mathcal{U}(u)}{\sqrt{v-u}}\, e_{\mu\nu}\exp\Big({i\phi(u)}\Big)\ .
\end{equation}

The back-reaction on the background (contained in a specific modification of the metric functions $\gamma$ and $\psi$)
is given by the following equations, cf. (\ref{2.4.6}),
\begin{eqnarray}
(v-u)\,\psi^2_{,u}+\gamma_{,u}&=&-\textstyle{\frac{1}{16}}(v-u)\,\mathcal{A}^2\dot{\phi}^2\ ,\label{r1}\\
(v-u)\,\psi^2_{,v}-\gamma_{,v}&=&0\ ,\label{r2}\\
\psi_{,uv}-\frac{1}{2(v-u)}(\psi_{,v}-\psi_{,u})&=&0\label{r3}\ .
\end{eqnarray}
Interestingly, this set of equations is \emph{consistent}: by differentiating equation (\ref{r1}) with respect to $v$,
equation (\ref{r2}) with respect to $u$, and combining them, one obtains (\ref{r3}) provided the amplitude
$\mathcal{A}(u,v)$ satisfies the equation
\begin{equation}
\Big((v-u)\,\mathcal{A}^2\Big)_{,v}=0\ .
\end{equation}
However, this is automatically satisfied for the amplitude (\ref{amplit}).
It is thus quite simple to introduce gravitational waves in the WKB approximation
into the cylindrical spacetimes (\ref{cylmetr}). If the functions $\gamma$ and $\psi$ representing
the background are solutions of the vacuum equations
[i.e. (\ref{r1})-(\ref{r3}) with a vanishing right-hand side of (\ref{r1})] then for introducing
high-frequency gravitational waves it is sufficient \emph{just to alter the
function} $\gamma$ as
\begin{equation}
\gamma(u,v)\rightarrow\gamma(u,v)+\tilde{\gamma}(u)\ ,\label{change}
\end{equation}
where
\begin{equation}
\frac{\partial\tilde{\gamma}(u)}{\partial u}=-\textstyle{\frac{1}{16}}\,\mathcal{U}^2\dot{\phi}^2\ .
\label{noveg}
\end{equation}
In particular, when ${\psi=0=\gamma}$ the background (\ref{cylmetrp}) is a flat Minkowski space. By assuming
non-trivial $\tilde{\gamma}$ we obtain Petrov type~$N$ spacetime with high-frequency gravitational
waves which have cylindrical wave-fronts. In a general case this
perturbation is propagating in the background which is the Einstein-Rosen cylindrical wave of
Petrov type~I. The effect on background  is given by the relation (\ref{noveg})
where  ${\mathcal{U}(u)=O(\epsilon)}$ is an arbitrary amplitude function.

The above described perturbations depend on the null ``retarded'' coordinate $u$ so that the
high-frequency gravitational waves are \emph{outgoing} ($\rho$ is
growing with $t$, on a fixed $u$). However, since the background metric (\ref{cylmetr}) is invariant
with respect to interchanging $u$ with $v$, it is straightforward to consider also
\emph{ingoing} perturbations by assuming the phase to depend on
the ``advanced coordinate'' $v$, namely
\begin{equation}
h_{\mu\nu}=\frac{\mathcal{V}(v)}{\sqrt{u-v}}\, e_{\mu\nu}\exp\Big({i\phi(v)}\Big)\ .
\end{equation}
Then the term proportional to $\mathcal{A}^2\dot{\phi}^2$ will appear
on the right-hand side  of equation (\ref{r2}) instead of (\ref{r1}). This results in
an interesting possibility to \emph{introduce ingoing high-frequency
gravitational cylindrical waves into the background of outgoing
Einstein-Rosen waves} just by assuming $\tilde{\gamma}(v)$ in (\ref{change}) such that
\begin{equation}
\frac{\partial\tilde{\gamma}(v)}{\partial v}=+\textstyle{\frac{1}{16}}\,\mathcal{V}^2\dot{\phi}^2\ ,
\label{novegn}
\end{equation}
or vice versa.

Moreover, all the above results can further be extended to a class of generalized
Einstein-Rosen (diagonal) metrics \cite{carmeli-charach,carmelichar} which describe  $G_2$
\emph{inhomogeneous cosmological models},
\begin{equation}
ds^{2}=e^{2\gamma-2\psi}(-\d t^{2}+\d\rho^{2})+ e^{2\psi}\d z^{2}+t^{2}e^{-2\psi}\d\varphi^{2}\ .
\label{cosm}
\end{equation}
If the three-dimensional spacelike hypersurfaces are compact, the corresponding model is
the famous Gowdy universe with the topology of three-torus \cite{gowdy,carmeli-charach}.
In the double null coordinates just one component of the metric is now different from (\ref{cylmetr}),
namely $g_{\varphi\varphi}=\frac{1}{2}(v+u)^{2}e^{-2\psi(u,v)}$.
The only modification of the above results (in the double null coordinates) consists of replacing
the factor $(v-u)$ with $(v+u)$, and each derivative with respect to $u$ changing  sign
(e.g. $\gamma_{,u}\rightarrow -\gamma_{,u}$ or $\psi_{,uv}\rightarrow -\psi_{,uv}$).
High-frequency gravitational waves in inhomogeneous cosmologies of the form (\ref{cosm})
can thus easily be constructed.

\subsection{Expanding waves}
Finally, we assume that the background is an expanding   Robinson-Trautman spacetime.
The metric (generally of the Petrov type~$II$)  in the standard coordinates has the
form, see e.g. \cite{rob-traut1,rob-traut2,kramerbook,BicPod99I},
\begin{equation}\label{rob}
ds^2=-\left(K-2r(\ln{\cal P})_{,u}-2\frac{m}{r}-\frac{\Lambda}{3}r^2\right)\d u^2-2\d u\d r
+\frac{r^{2}}{{\cal P}^{2}}(\d\eta^{2}+\d\xi^{2})\ ,
\end{equation}
where  $K=\Delta(\ln{\cal P})$,  $\Delta\equiv {\cal P}^{2}(\frac{\partial^{2}}{\partial\eta^{2}}
+\frac{\partial^{2}}{\partial\xi^{2}})$,
and $m(u)$.
When  ${\cal P}(u,\eta,\xi)$  satisfies the Robinson-Trautman equation
$\Delta K+12\,m\,(\ln{\cal P})_{,u}-4m_{,u}=0$, the metric (\ref{rob}) is an exact
vacuum solution of the Einstein equations.

In view of the existence of privileged congruence of null geodesics generated by ${\partial_r}$
we introduce the phase $\phi=\phi(u)$ and the wave vector
$k_{\mu}=(\dot{\phi},0,0,0)$ of high-frequency gravitational waves.
We again assume the WKB form (\ref{x1.33b}) of the solution. Applying the equations (\ref{2.4.4b}) we
obtain
\begin{eqnarray}
\mathcal{A}&=&\frac{1}{r}\,U(u,\eta,\xi)\ ,\nonumber\\
e^{+}_{\mu\nu}&=&\frac{1}{\sqrt{2}}\frac{r^{2}}{{\cal P}^{2}}
\left(
\begin{array}{cccc}
0&0&0&0\\
0&0&0&0\\
0&0&1&0\\
0&0&0&-1
\end{array}
\right)\ ,\label{RTvlny}\\
e^{\times}_{\mu\nu}&=&\frac{1}{\sqrt{2}}\frac{r^{2}}{{\cal P}^{2}}
\left(
\begin{array}{cccc}
0&0&0&0\\
0&0&0&0\\
0&0&0&1\\
0&0&1&0
\end{array}
\right)\ .\nonumber
\end{eqnarray}
A general solution has the form
$\>h_{\mu\nu}=r^{-1}U(u,\eta,\xi)\,e_{\mu\nu}\exp\Big({i\phi(u)}\Big)$,
where $U(u,\eta,\xi)$ and $\phi(u)$ are arbitrary functions, and
$e_{\mu\nu}=a\,e^{+}_{\mu\nu}+b\,e^{\times}_{\mu\nu}$ with
$a^{2}(u,\eta,\xi)+b^{2}(u,\eta,\xi)=1$. Introducing the
amplitudes $U^{+}=a\,U$, $U^{\times}=b\,U$ for both polarizations, we can write the solution as
\begin{equation}\label{robres}
h_{\mu\nu}=\frac{1}{r}\left[U^{+}e^{+}_{\mu\nu}+U^{\times}e^{\times}_{\mu\nu}\right]
 \exp\Big({i\phi(u)}\Big)\ .
\end{equation}
If the wave-surfaces $r=const.,u=const.$ with the metric $dl^{2}={\cal P}^{-2}(\d\eta^{2}+\d\xi^{2})$
are homeomorfic to  $\mathcal{S}^{2}$, the waves can be interpreted as  ``spherical''.
In the asymptotic region  $r\rightarrow \infty$ such solutions locally approach plane waves
\cite{hogan}.

The reaction of the waves on  background is determined by the equations (\ref{my.5}) and (\ref{2.4.6})
with $T_{\mu\nu}^{(0)}=-\frac{1}{8\pi}{\Lambda}\gamma_{\mu\nu}$.
From the only nontrivial component we immediately obtain the following equation
\begin{equation}\label{back}
-\frac{\partial m}{\partial u}+3\, m\,(\ln{\cal P})_{,u}+{\textstyle \frac{1}{4}}\Delta K
 ={\textstyle \frac{1}{16}}\Big[(U^{+})^{2 }+(U^{\times})^{2}\Big]{\dot{\phi}}^{2}\ ,
\end{equation}
where $m(u)$, $\phi(u)$, whereas the remaining functions depend on coordinates
$\{u,\eta,\xi\}$. Notice that this is \emph{independent} of the cosmological
constant $\Lambda$.

The expressions (\ref{robres}),(\ref{back})  agree with results obtained by
MacCallum and Taub \cite{maccallumtaub} or recently by Hogan and Futamase \cite{hogan}
who used Burnett's technique \cite{burnett}.
Our results, which were derived by a straightforward approach, are slightly more general
because they are not restricted to a constant frequency ${\dot\phi=const}$. Particular  subcase of the Vaidya
metric has already been studied before by Isaacson \cite{isaac} and elsewhere \cite{choquet2}.

\section{Conclusions}
The Isaacson approach to study high-frequency perturbations of Einstein's
equations was briefly reviewed and compared with the standard weak-field limit.
In our contribution we generalized Isaacson's method to include non-vacuum
spacetimes, in particular an electromagnetic field and/or a non-vanishing
value of the cosmological constant $\Lambda$. Then we explicitly analyzed
possible high-frequency gravitational waves in three large families of
background universes, namely non-expanding spacetimes of the Kundt type,
cylindrical Einstein-Rosen waves and related inhomogeneous cosmological
models (such as the Gowdy universe), and the Robinson-Trautman
expanding spacetimes. These backgrounds are of various Petrov
types. For example, high-frequency gravitational waves can be
introduced into electrovacuum conformally flat  Bertotti-Robinson space,
type~$D$ Nariai and Pleba\'nski-Hacyan spaces, their type~$N$ and type~$II$
generalizations, or into algebraically general Einstein-Rosen universes.

For construction of high-frequency gravitational perturbations we have
employed the fact that all these spacetimes admit a non-twisting
congruence of null geodesics. The corresponding tangent vectors $k^{\mu}$
are hypersurface orthogonal so that there exists a phase function $\phi$  which satisfies
${\phi_{,\mu}=k_{\mu}}$. The last equation in (\ref{2.4.4b}) can be put into the form
${\textstyle{\frac{d}{dl}}(\ln\mathcal{A})=-\Theta}$,
where $l$ is the affine parameter, and ${\Theta=\frac{1}{2}k^{\mu}_{\>;\mu}}$
is the expansion of the null congruence. This determines the
behaviour of the amplitude $\mathcal{A}$ in the above spacetimes
(\ref{tenz}), (\ref{amplit}), (\ref{RTvlny}). The remaining
equations (\ref{2.4.4b}) enables one to deduce the polarization
tensors.

It has been also crucial that all the classes of spacetimes discussed admit
\emph{exact} solutions with the energy-momentum tensor of  pure
radiation, i.e., ${G_{\mu\nu}-8\pi\,T_{\mu\nu}=
\textstyle{\frac{1}{8}}\,\mathcal{A}^{2}k_{\mu}k_{\nu}}$,
where $T_{\mu\nu}$ is either constant (representing the
cosmological constant) or it describes an electromagnetic field.
The relation between high-frequency perturbations and exact
radiative solutions of Einstein's equations in each class is thus natural.
In particular, it is possible to determine explicitly
the reaction of the background on the presence of high-frequency
gravitational waves.

\section*{Acknowledgements}
The work was supported in part by the grants GA\v{C}R 202/02/0735 and GAUK 166/2003
of the Czech Republic and the Charles University in Prague.


\begin{thebibliography}{45}
\bibitem{isaac}
Isaacson, R.~A. (1968).
{\em Gravitational Radiation in the Limit of High Frequency I. and II.},
Phys. Rev. {\bf 166}, 1263-1280.

\bibitem{einstein1}
Einstein, A. (1916).
{\em N\" aherungsweise Integration der Feldgleichungen der Gravitation},
Preuss. Akad. Wiss. Sitz. {\bf 1}, 688-696.

\bibitem{einstein2}
Einstein, A.  (1918).
{\em \" Uber Gravitationswellen},
Preuss. Akad. Wiss. Sitz. {\bf 1}, 154-167.

\bibitem{MTW}
Misner, C.~W., Thorne, K.~S., and Wheeler J.~A. (1973).
{\em Gravitation} (W.~H.~Freeman: San Francisco).

\bibitem{Wheeler}
Wheeler, J.~A. (1962).
{\em Geometrodynamics} (Academic Press: New York).

\bibitem{bh}
Brill, D.~R., and Hartle, J.~B. (1964).
{\em Method of the Self-Consistent Field in General Relativity and its Application to the Gravitational Geon},
Phys. Rev. {\bf 135}, B271-B278.

\bibitem{choquet1}
Choquet-Bruhat, Y. (1968).
{\em Construction de solutions radiatives approch\'ees des \'equations d'Einstein},
Rend. Acc. Lincei {\bf 44}, 345-348.

\bibitem{choquet2}
Choquet-Bruhat, Y. (1968).
{\em Construction de solutions radiatives approch\'ees des \'equations d'Einstein},
Commun. Math. Phys. {\bf 12}, 16-35.

\bibitem{maccallumtaub}
MacCallum, M.~A.~H., and Taub, A.~H. (1973).
{\em The Averaged Lagrangian and High-Frequency Gravitational Waves},
Commun. Math. Phys. {\bf 30}, 153-169.

\bibitem{taub}
Taub, A.~H. (1980).
{\em High-Frequency Gravitational Waves, Two-Timing, and Avaraged Lagrangians},
in {\em General Relativity and Gravitation}, Vol. 1, ed. A.~Held
(Plenum Press: New York), p. 539-555.

\bibitem{araujo1}
Araujo, M.~E. (1986).
{\em On the Assumptions Made in Treating the Gravitational Wave Problem by the High-Frequency Approximation},
Gen. Rel. Grav. {\bf 18}, 219-233.

\bibitem{araujo2}
Araujo, M.~E. (1989).
{\em Lagrangian Methods and Nonlinear High-Frequency Gravitational Waves},
Gen. Rel. Grav. {\bf 21}, 323-348.

\bibitem{elster}
Elster, T. (1981).
{\em Propagation of High-Frequency Gravitational Waves in Vacuum: Nonlinear Effects},
Gen. Rel. Grav. {\bf 13}, 731-745.

\bibitem{burnett}
Burnett, G.~A. (1989).
{\em The high-frequency limit in general relativity},
J. Math. Phys. {\bf 30}, 90-96.

\bibitem{taub2}
Taub, A.~H. (1976).
{\em High Frequency Gravitational Radiation in Kerr-Schild Space-Times},
Commun. Math. Phys. {\bf 47}, 185-196.

\bibitem{hogan}
Hogan, P.~A., and Futamase, T. (1993).
{\em Some high-frequency spherical gravity waves},
J. Math. Phys. {\bf 34}, 154-169.

\bibitem{brinkmann}
Brinkmann, H.~W. (1925).
{\em On Riemann Spaces conformal to Euclidean spaces},
Proc. Natl. Acad. Sci. U.S. {\bf 9}, 1.

\bibitem{BPR}
Bondi, H., Pirani, F.~A.~E., and Robinson I. (1959).
{\em Gravitational waves in general relativity, III. Exact plane waves},
Proc. Roy. Soc. Lond. {\bf A251}, 519-533.

\bibitem{kundt}
Kundt, W. (1961).
{\em The Plane-fronted Gravitational Waves},
Z.~Phys. {\bf 163}, 77-86.

\bibitem{ehlers}
Ehlers, J., and Kundt, K. (1962).
{\em Exact solutions of the gravitational field equations},
in {\em Gravitation: an Introduction to Current Research}, ed. L~Witten
(J.~Wiley~\& Sons: New York), p. 49-101.

\bibitem{einst-ros}
Einstein, A., and Rosen, N. (1937).
{\em On Gravitational Waves},
Journ. Franklin. Inst. {\bf 223}, 43-45.

\bibitem{rob-traut1}
Robinson, I., and Trautman, A. (1960).
{\em Spherical Gravitational Waves},
Phys. Rev. Lett. {\bf 4}, 431-432.

\bibitem{rob-traut2}
Robinson, I., and Trautman, A. (1962).
{\em Some spherical gravitational waves in general relativity},
Proc. Roy. Soc. Lond. {\bf A265}, 463-473.

\bibitem{BonSwa64}
Bonnor, W.~B., and Swaminarayan, N.~S. (1964).
{\em An exact solution for uniformly accelerated particles in general relativity},
Z.~Phys. {\bf 177}, 240-256.

\bibitem{Bic68}
Bi\v{c}\'ak, J. (1968).
{\em Gravitational radiation from uniformly accelerated particles in general relativity},
Proc. Roy. Soc. A {\bf 302}, 201-224.

\bibitem{KinWal70}
Kinnersley, W., and Walker, M. (1970).
{\em Uniformly accelerating charged mass in general relativity},
Phys. Rev. D {\bf 2}, 1359-1370.

\bibitem{gowdy}
Gowdy, R.~H. (1971).
{\em Gravitational waves in closed universes},
Phys. Rev. Lett. {\bf 27}, 826-829.

\bibitem{kramerbook}
Kramer, D., Stephani, H., MacCallum, M.~A.~H., and Herlt, E. (1980).
{\em Exact Solutions of  Einstein's Field Equations}
(Cambridge University Press: Cambridge).

\bibitem{carmeli-charach}
Carmeli M., Charach Ch., and Malin S.  (1981).
{\em Survey of cosmological models with gravitational, scalar and electromagnetic waves},
 Phys. Rep. {\bf 76}, 79-156.

\bibitem{BicSch89b}
Bi\v{c}\'ak, J., and Schmidt, B.~G. (1989).
{\em Asymptotically flat radiative space-times with boost-rotation symmetry: The general structure},
Phys. Rev. D {\bf 40}, 1827-1853.

\bibitem{BoGrMa94}
Bonnor, W.~B., Griffiths, J.~B., and MacCallum, M.~A.~H. (1994).
{\em Physical Interpretation of Vacuum Solutions of Einstein's Equations. Part II. Time-dependent Solutions},
Gen. Rel. Grav. {\bf 26}, 687-729.

\bibitem{Bic}
Bi\v{c}\'ak, J. (2000).
{\em Selected Solutions of Einstein's Field Equations: Their Role in General Relativity and Astrophysics},
in {\em Einstein's Field Equations and Their Physical Implications}, ed. B~G~Schmidt
(Springer Verlag: Berlin), p. 1-126.

\bibitem{feinstein88}
Feinstein, A. (1988).
{\em Late-time behavior of primordial gravitational waves in expanding universe},
Gen. Rel. Grav. {\bf 20}, 183-190.

\bibitem{WMAP}
Bennett, C.~L. et~al. (2003).
{\em First year Wilkinson Microwave Anisotropy Probe (WMAP) observations: preliminary maps and basic results},
Astrophys. J., to appear [astro-ph/0302207].

\bibitem{efroimsky}
Efroimsky, M. (1992).
{\em Gravity waves in vacuum and in media},
Class. Quantum Gravity {\bf 9}, 2601-2614

\bibitem{PodOrt03}
Podolsk\'y, J.,  and Ortaggio, M. (2003).
{\em Explicit Kundt type II and N solutions as gravitational waves in various type D and O universes},
Class. Quantum Grav. {\bf 20}, 1685-1701.

\bibitem{OzsRobRoz85}
Ozsv\'ath, I., Robinson, I., and R\'ozga, K. (1985).
{\em Plane-fronted gravitational and electromagnetic waves in spaces with cosmological constant},
J. Math. Phys. {\bf 26}, 1755-1761.

\bibitem{Siklos85}
Siklos, S.~T.~C. (1985).
{\em Lobatchevski Plane Gravitational Waves},
in {\em Galaxies, Axisymmetric Systems and Relativity}, ed.
M~A~H MacCallum (Cambridge University Press: Cambridge), p. 247-274.

\bibitem{BicPod99I}
Bi\v{c}\'ak, J., and Podolsk\'y, J. (1999).
{\em Gravitational waves in vacuum spacetimes with cosmological constant.
I. Classification and geometrical properties of nontwisting type N solutions},
J. Math. Phys. {\bf 40}, 4495-4505.

\bibitem{Nariai51}
Nariai, H. (1951).
{\em On a new cosmological colution of Einstein's field equations of gravitation},
Sci. Rep. T\^{o}hoku Univ. {\bf 35}, 62-67.

\bibitem{Bertotti59}
Bertotti, B. (1959).
{\em Uniform electromagnetic field in the theory of general relativity},
Phys. Rev. {\bf 116}, 1331-1333.

\bibitem{Robinson59}
Robinson, I. (1959).
{\em A Solution of Maxwell-Einstein Equations},
Bull. Acad. Polon. {\bf 7}, 351-352.

\bibitem{PlebHac79}
Pleba\'nski, J.~F., and Hacyan, S. (1979).
{\em Some exceptional electrovac type~D matrics with cosmological constant},
J. Math. Phys. {\bf 20}, 1004-1010.

\bibitem{OrtPod02}
Ortaggio, M., and Podolsk\'y, J. (2002).
{\em Impulsive waves in electrovac direct product spacetimes with $\Lambda$},
Class. Quantum Grav. {\bf 19}, 5221-5227.

\bibitem{carmelichar} Carmeli, M., and Charach, Ch. (1984).
{\em The Einstein-Rosen gravitational waves and cosmology},
 Found. Phys. {\bf 14}, 963-986.

\end{thebibliography}
\end{document}